\documentstyle[12pt]{article}
\input epsf
\epsfverbosetrue
\newcommand{\be}{\begin{equation}}
\newcommand{\ee}{\end{equation}}

\begin{document}

\begin{flushright}
NORDITA 98/79 \\
LTH -- 449 \\
\today
\end{flushright}

\begin{centering}

{\Large{\bf Two heavy-light mesons on a lattice}}

\vspace{0.5cm}
{\em UKQCD Collaboration}

\vspace{0.25cm}
C. Michael\footnotemark[1]$^,$\footnotemark[2] and
P. Pennanen\footnotemark[3]$^,$\footnotemark[4]

\vspace{0.125cm}

{\em $^1$Theoretical Physics Division, Dept. of Math. Sciences,
University of Liverpool, Liverpool, UK.}

{\em $^3$Nordita, Blegdamsvej 17, DK-2100 Copenhagen \O, Denmark}

\setcounter{footnote}{2}
 \footnotetext{E-mail: {\tt cmi@liv.ac.uk}}
\setcounter{footnote}{4}
\footnotetext{{\tt petrus@hip.fi}}
\renewcommand{\thefootnote}{\arabic{footnote}}

\vspace{0.75cm}

{\bf Abstract}\\  The potential between two heavy-light mesons as a
function of the heavy quark separation is calculated in quenched SU(3) 
lattice QCD. We study the case of heavy-light mesons
with a static heavy quark and light quarks  of mass close to the strange
quark mass.   We explore the case of  light quarks with the same and
with different flavours, classified according to the light quark
isospin. We evaluate the appropriate light quark exchange contributions
and explore  the spin-dependence of the interaction. Comparison 
is made with meson exchange. 

\vspace{0.25cm}
\end{centering}

\noindent


\vspace{1cm}

\section{Introduction}

The progress in lattice QCD has so far been mainly restricted to systems
of three quarks or less. However, there is also considerable 
interest in obtaining predictions from first principles for multi-quark 
systems which can be decomposed into more than
one colour singlet. In addition to the complicated cases of nuclei, simple
multi-quark systems have been proposed to exist as bound 
states~\cite{vol:76,jaf:76,gut:79}. Four quarks forming colour 
singlets or as bound states of two mesons are candidates for particles lying 
close to meson-antimeson threshold, such as $a_0(980),\ f_0(980)$ 
($K\bar{K}$), $f_0(1500),\ f_2(1500)$ ($\omega\omega,\ \rho\rho$), 
$f_J(1710)$ ($K^*\bar{K}^*$), $\psi(4040)$ ($D^*\bar{D}^*$), 
$\Upsilon(10580)$ ($B^*\bar{B}^*$)~\cite{pdg:98}. 

Systems with heavy quarks should be more easily bound provided the
potential is attractive, since the repulsive kinetic energy of the quarks
is smaller, while the attractive two-body potential remains the same. In
so-called deuson  models~\cite{tor:91} the long-range potential
between two mesons  comes from one-pion exchange, suggesting that 
meson-meson systems are significantly less bound than meson-antimeson
systems. Other models used for four-quark systems include string-flip
potential  models (see Ref.~\cite{boy:96} for a review), bag
models~\cite{jaf:76}, and a model-independent approach~\cite{lip:86}.
Four-quark states  with two heavy quarks have been
predicted to be stable~\cite{zou:86}.  Most models give stability
for systems where the heavy quarks have the  $b$ mass, but long range
forces might push the required heavy-to-light mass  ratio down so that
$cc\bar{q}\bar{q}$ states would be bound as well.

 Static four-quark systems~\cite{glpm:96} (and  references therein)  
have been previously studied  for a set of geometries representative of
the general case and a model was constructed that reproduces one
hundred ground and excited state energies with four independent
parameters~\cite{gp}. The model is based on ground- and excited state
two-body potentials and multi-quark  interaction terms. The results
show that, in a two-body potential approach to understanding
multi-quark interaction, the effect from gluonic excitations is needed,
and their relative contribution to the binding becomes more  important
(even dominant) at larger distances. Flux distributions corresponding
to the binding energies of four static quarks are studied in
Ref.~\cite{pen:98}.

Moving on to more realistic systems, we now study in detail the
potential between  two heavy-light mesons. Exploratory studies of
two-meson systems have been made for the cross  diagram only (Fig.~1
below) for SU(3) colour~\cite{BBold} and for both  diagrams  in 
Refs.~\cite{BBcanada},~\cite{mih:97} for SU(2), SU(3) colour 
respectively.

 We take the mass of one quark in  each meson to be heavy -- the
prototype being the B meson. This is in the spirit of  the  heavy quark
effective theory approach which describes the leading term (the static 
limit) and  the corrections of higher orders in $1/m_Q$. In the  static
approximation for the heavy quarks, the pseudoscalar B meson and the
vector B$^*$ meson  will be degenerate -- whereas they are split by 46
MeV experimentally. Since we shall often have occasion to treat this 
degenerate set together we describe this case as the ${\cal B}$ meson.
In  analogy to the Born-Oppenheimer approximation, we will then discuss
the  potential energy between static  ${\cal B}$ mesons. 

For the light quarks, we use the full relativistic description with a
fermion action which is  the $O(a^2)$ improved Sheikholeslami-Wohlert
clover action with a  tadpole-improved coefficient. We should in
principle evaluate the interaction for several light quarks  masses and
then extrapolate to the physical values. In this preliminary  study, we
fix the light quark mass at around the strange mass. We do  however
consider the case of two flavours of quark -- so allowing a  discussion
of different isospin states. The main reason why this  study is
difficult to perform on a lattice is that the light quark propagators
are needed from many different sources. To achieve this  we make use of
the technique of evaluating the light quark propagators  as stochastic
estimates~\cite{div:96} using maximal variance reduction introduced in
Ref.~\cite{mic:98}. 

Quenched lattices are used with SU(3)
colour and static  heavy quarks  with light quarks of  approximately the
strange quark mass.  Preliminary versions of this work have
appeared~\cite{conf}.  Here the isospin  and spin degrees of freedom
are discussed in detail. We  compare our results for small separation
$R$ with the known spectrum of baryons with one heavy quark ($\Lambda_b$
 and $\Sigma_b$). This will enable us to discover if a heavy diquark  is
a good description. Note that this  link which  we find to baryons at
small separation $R$ cannot be explored using  SU(2) of colour.  We
also compare our results with the expectations  of meson exchange. We
find that at larger $R$, this is a useful guide  to the interaction 
strength and, for pion exchange, we are able to make a quantitative
comparison. We comment on the agreement with other models, one of them
being the potential model for static systems applied in this more
dynamic case~\cite{gre:99}. 

\section{${\cal B}$${\cal B}$ interactions in the static approximation}

We take the mass of one quark in  each meson to be very heavy -- the
prototype being the B meson.  The static limit is then the leading term
in the heavy quark effective theory for a heavy quark  of zero velocity
and there will be   corrections of higher orders in $1/m_Q$ where $m_Q$
is the heavy quark mass.  In the limit of a static heavy quark, the
heavy quark  spin is uncoupled since the relevant magnetic moment
vanishes which implies  that  the pseudoscalar B meson and the vector
B$^*$ meson  will be degenerate. This is a reasonable  approximation  
since they are split by 46 MeV experimentally, which is less than 1\% of the
mass of the mesons. Since we shall often have
occasion to treat these  two mesonic states as if they were degenerate,
we describe them collectively as the ${\cal B}$ meson.  Because of the
insensitivity to the heavy quark spin,  it is then  appropriate to
classify  these degenerate ${\cal B}$ meson states by the light quark
spin: so there are only two  independent spin states. 
 The  system of two heavy-light mesons at spatial separation $R$ will 
be referred to as the ${\cal BB}$ system. With
both heavy-light mesons static, this ${\cal BB}$ system is
described  by the spin states of the two light quarks in the two mesons.
Thus there are four possible states  and we need to classify the
interaction in terms of these spin states.

This situation is very similar to that of the hydrogen molecule in the
Born-Oppenheimer approximation -- with, however,  the additional
possibility that the two `electrons' can have different properties.
  Another similarity is with  the potential between   quarks which has a
central component and then scalar and tensor spin-dependent
contributions. 

Each ${\cal B}$ meson will have a light quark flavour assignment. For
the ${\cal BB}$  system, it will be appropriate to classify these
states according to their symmetry under interchange of the light quark
flavours. For  identical flavours (eg. ${s} s$ or ${u} u$), we have
symmetry  under interchange, whereas  for non-identical flavours (eg.
${s}u$  or ${d} u$), we may have either symmetry or antisymmetry. For 
two light quarks, it is convenient to classify the states according to 
isospin as $I=1$ (with $uu$, $ud+du$ and $dd$) or $I=0$ (with 
$ud-du$).

 We now present a discussion of the possible states of two ${\cal B}$
mesons. As a guide we show in Table~1 the states for the case of an 
S-wave ${\cal BB}$ system in the limit of  static ${\cal B}$ mesons.  We
must have overall symmetry of the wavefunction  under interchange and,
assuming symmetry for spatial interchange, the  flavour, total light
quark spin ($S_q$) and total heavy quark spin ($S_b$) must be  combined to 
achieve  this.  
 Thus in the limit of an isotropic spatial wavefunction, there will be 
the four different ground state levels of the ${\cal BB}$ system as 
shown in Table~\ref{table1} since the three states with different $J^P$
but the same light quark isospin $I_q$ and spin  $S_q$ will be
degenerate in the static limit. We will label these states by $I_q,\ S_q$
for subsequent discussion. We also show which physical B and B$^*$
mesons couple to these states. 
 This table can also be extended to $L \ne 0$ levels. In particular, we
shall later see that a tensor interaction may be present,  in which case
the $S_q=1$ ground states will show an admixture of $L=0$  and of $L=2$.

\begin{table}[hbt]

\caption{Allowed ${\cal BB}$ states with $L=0$}
\begin{center}
\begin{tabular}{cccllll}
 $I_q$ & $S_q$ &  $S_b$ &   $J^{P}$ &  BB & BB$^*$ &  B$^*$B$^*$ \\
\hline
 1   & 1  &  1  &  0$^+$  & Yes   &       & Yes\\
 1   & 1  &  1  &  1$^+$  &    & Yes      &\\
 1   & 1  &  1  &  2$^+$  &    &         & Yes\\
 1   & 0  &  0  &  0$^+$  & Yes &         & Yes\\
 0   & 1  &  0  &  1$^+$  &    & Yes      & Yes\\
 0   & 0  &  1  &  1$^+$  &    & Yes      & Yes\\

\end{tabular}
 \label{table1}
\end{center}
\end{table}

 When $R=0$, the situation is special since the colour of the  two
static quarks can be combined. This net colour can be in a  anti-triplet
(antisymmetric under particle exchange) or a sextet (symmetric under
particle exchange). The former case is just that of the static baryons.
This equivalence implies that the $I_q=1,\ S_q=1$ state  will have the
same  light quark structure as the $\Sigma_b$ baryon, while the $I_q=0,\
S_q=0$ state will be as  the $\Lambda_b$ baryon. The other two allowed
${\cal BB}$ states at $R=0$ correspond  to a static sextet source.

In a Born-Oppenheimer treatment of the ${\cal BB}$ system, we will 
need to consider the potential energy for the ${\cal B}$ mesons at rest
at  separation $R$.  This  ${\cal BB}$ system can be classified under
rotations about the separation axis,  here taken as the $z$ axis, and
under interchange of the two mesons. Taking the $z$ axis to quantise
the light quark spin, we have states  with $J_z=\pm 1$, namely
$|++\rangle,\ |--\rangle$ and with  $J_z=0$, namely $|+-\rangle \pm
|-+\rangle$. Since $J_z$ is conserved, we  can discuss the interaction
energy in terms of a triplet state $E_T$  corresponding to the
$J_z=\pm1$ cases, and then the $J_z=0$ sector can be  described by a
singlet state with energy $E_S$ between initial and final  states
$(|+-\rangle - |-+\rangle)$ and by another triplet state $E_{T}'$ for 
initial and final  states $(|+-\rangle + |-+\rangle)$.
 These three energies can be related to a more conventional treatment
using a central, spin-dependent  and tensor potential. We shall instead
 mainly focus on the singet, $E_S$, and triplet averaged over
orientations, ($2E_T+E_T')/3$.
 Since the heavy quark  spin does not interact, the symmetric and
antisymmetric combinations constructed from the heavy quark spin will
allow any overall symmetry under interchange  for the overall spin
assignment.

 In our actual numerical calculation, we use a relativistic treatment of
the  light quark spin but in the context of a static heavy quark with
Dirac propagator structure $(1+\gamma_4)$. This enables us to simplify
the  Dirac $\gamma$-matrix algebra between initial and final B mesons
(created  by $\bar{q} \gamma_5 b$) and B$^*$ mesons (created by $
\bar{q} \gamma_i b$). This approach also leads to three independent
observables which we determine as 
 $$
  C_I= ((++) + (--)).((++) + (--))
 $$
 $$
  C_s(z)= ((++) - (--)).((++) - (--))
 $$
 $$
  C_s(x)=C_s(y)= (+-).(-+)  +  (-+).(+-)
 $$
 with notation $(13).(24)$ for $B_1 B_2 \to B_3 B_4$ with the sign of 
the light quark spin ($S_{1z}$, etc) given. 

 In practice the observable given above by $C_s(z)$ is also evaluated
with the spatial separation $R$ in  $x$ and $ y$ directions which gives
an equivalent method to obtain $C_s(x)$ and $C_s(y)$. By symmetry, the
latter two observables are  equal on average.
 Note that the BB\ $\to$\ BB correlation is given by $C_I$, whereas
BB$^*\ \to$\ B$^*$B is given by $C_s$. 

 It is not sufficient just to look at processes such as BB\ $\to$\ BB 
since, in the heavy quark limit, there will also be other channels  such
as BB\ $\to$\ B$^*$B$^*$ which are coupled. We then analyse the matrix 
of correlations between all such channels and find the basis that 
diagonalises it. This leads to certain linear combinations of 
correlations which describe these diagonal elements.  This  explicit
fermionic approach must reproduce the conclusions reached above by 
using the heavy quark limit.  The relationship between these approaches
is that, at large $t$, 
 \be
 C_I+C_s(z) \rightarrow e^{-E_T t}
 \ee
 \be 
 C_I-C_s(z)-C_s(x)-C_s(y) \rightarrow e^{-E_S t}
 \ee
 \be
 C_I-C_s(z)+C_s(x)+C_s(y) \rightarrow e^{-E_T' t}
 \ee
 It turns out that the same combinations (those given in the above
equations) occur  for both the case of symmetry under exchange of 
initial particles and for the case of antisymmetry. This can be
understood  in the heavy quark effective theory, as discussed above,
since the combinations are in terms  of the light quarks spins, leaving
the heavy quark spins to be combined  either in symmetric or
antisymmetric states.

 The structure of the correlations to be evaluated, in terms of the 
light quark propagator $G$ and the gauge product for the static line in 
the negative-going $t$-direction of $U$ is  then
 \be
  C_I(t)= \langle G^{ba}_{ii}(0,0;0,t)U^{ab}(0) \ 
          G^{dc}_{kk}(e_zR,0;e_zR,t)U^{cd}(e_zR) \rangle
 \ee
 where the colour indices $a,b,c,d$ and the Dirac indices $i,j,k,l$ are 
associated with vertices as in Fig.~1. The sum over Dirac indices is
only  from 1 to 2 since the heavy  quark has a spin projection factor.  
 These contributions can be evaluated for every choice of 
origin on a lattice which is translationally invariant.
 For the spin-dependent part 
with component $p$, we have
 \be
  C_s^p(t)= \langle G^{ba}_{ji}(0,0;0,t)\sigma^p_{ij}U^{ab}(0) \ 
          G^{dc}_{lk}\sigma^p_{kl}(e_zR,0;e_zR,t)U^{cd}(e_zR) \rangle
 \ee

 For the `cross' diagram the colour and spin sums are different, for 
example the contribution to $C_I$ is given by
 \be
  C_I(t)= -\langle G^{da}_{ki}(0,0;e_zR,t)U^{cd}(e_zR) \ 
          G^{bc}_{ik}(e_zR,0;0,t)U^{ab}(0) \rangle
 \ee
 where the negative sign comes from the Grassmannian nature of the 
fermions. For states symmetric under light quark interchange (eg. $I=1$), 
then the sum of uncrossed and crossed diagram is needed, where the 
above minus sign is incorporated into the crossed  diagram -- this plays the 
r\^ole of the Pauli principle. For states antisymmetric under light quark 
interchange (eg. $I=0$), the difference of uncrossed and crossed 
diagrams is needed.

\begin{figure}[h]
\vspace{5cm} 
\includegraphics{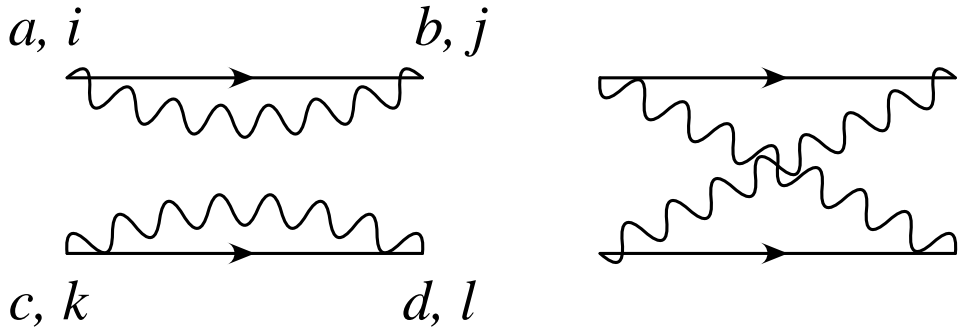} 
 \caption{Diagrams showing the interaction between two ${\cal B}$
mesons: the light  quarks are shown as wiggly lines.
 }
\label{bbdf}
\end{figure}

\section{Fermion formalism}

The diagrams we need to evaluate are illustrated in Fig.~1. We need 
light quark propagators from more that one source -- so the conventional 
approach of inverting from a single source is impractical.  One feasible
way forward is to use a stochastic inversion method  which allows the
evaluation of quark propagators from any site to any other site. The
stochastic method  has already been shown to be  more efficient than the
conventional inversion from one source  for mesons made of heavy-light
quarks~\cite{mic:98}, and it  does allow the flexibility  to evaluate
the  required combinations of correlations readily. For this reason it
allows  a thorough study of this area.

Stochastic propagators~\cite{mic:98,div:96} are one technique to
invert the fermionic matrix for the light quarks. They can be used in
place of light quark propagators calculated with the usual deterministic
algorithm.  The stochastic inversion is based on the relation:
 \begin{equation}
        G_{ij} =  {\cal M}_{ij}^{-1}=\frac1Z \int {\cal D}\phi\;
        ({\cal M}_{jk}\phi_k)^\ast \phi_i\; 
        \exp \left( -\phi_i^\ast ({\cal M}^\dagger {\cal M})_{ij}
        \phi_j \right) 
 \end{equation}  where, in our case, ${\cal M}$ is the improved
Wilson-Dirac fermionic operator and the indices $i,j,k$ represent
simultaneously the space-time coordinates, the spinor and colour
indices.  For every gauge configuration, an ensemble of independent fields
$\phi_i$ (we use 24 following~\cite{mic:98}) is generated with gaussian
probability:
 \begin{equation} 
P[\phi] =\frac1Z \exp \left(
-\phi_i^\ast ({\cal M}^\dagger {\cal M})_{ij} \phi_j \right) 
 \end{equation} 
All light propagators are computed as averages over the
pseudo-fermionic samples:
\begin{equation}
        G_{ij} = 
        \left\{\begin{array}{l}
               \langle ({\cal M}\phi)_j^\ast \phi_i \rangle \\
               \mathrm{or} \\
               \gamma_5 \langle \phi_j^\ast ({\cal M}\phi)_i \rangle\gamma_5 
               \end{array}
        \right. \label{eq:stock} 
 \end{equation} 
 where the two expressions are related by $G_{ij} = \gamma_5
G_{ji}^\dagger \gamma_5$.  Moreover, the maximal variance reduction
method is applied in order to minimise the statistical
noise~\cite{mic:98}. The maximal variance reduction method involves
dividing the lattice  into two boxes ($0<t<T/2$ and $T/2<t<T$) and
solving the equation of motion numerically within each box, keeping the
pseudo-fermion field $\phi$ on the boundary fixed.  According to the
maximal reduction method, the fields which enter the correlation
functions must be either the original fields $\phi$ or solutions of the
equation of motion in disconnected regions.  The stochastic propagator
is therefore defined from each point in one box to every point in the
other box or on the boundary. For more than one propagator from one 
box to the other, we need to use different stochastic samples for each. 
This is completely analogous to the technique used to discuss the 
$\Lambda_b$ meson~\cite{mic:98}.

The numerical analysis used 24 stochastic samples  on each of 60
quenched gauge configurations, generated~\cite{mic:98} on a $12^3 \times
24$ lattice at $\beta=5.7$, corresponding to $a^{-1}=1.10$ GeV.  With
improved clover coefficient $C_{SW}=1.57$, we use a value of
$\kappa_1=0.14077$ which corresponds to a bare mass of the light quark
around the strange mass  and gives a pseudoscalar to vector mass ratio
of 0.650(7). The chiral limit corresponds to
$\kappa_c=0.14351$~\cite{hugh97}. Error estimates come from bootstrap 
over the gauge configurations.  
 We also used 20 quenched gauge configurations on a $16^3 \times 24$ 
lattice to check finite size effects with the same parameters as above.
Allowing for the self-averaging effect of the larger  spatial volume,
this data set has similar weight to that at the  smaller volume.

In smearing the hadronic interpolating operators, spatial fuzzed links
are used. Following the prescription in~\cite{mic:98, michael95}, the
fuzzed links are defined iteratively as:
 \begin{equation}
        U_{\mathrm{new}} = {\mathcal P} \left(f U_{\mathrm{old}} 
          + \sum_{i=1}^4 U_{\mathrm{bend},i} \right) 
 \end{equation}
 where ${\mathcal P}$ is a projector over SU(3), and
$U_{\mathrm{bend},i}$ are the staples attached to the link in the
spatial directions. Two iterations of fuzzing with $f=2.5$ are used and
then  the fuzzed links of length one are used. The fuzzed fermionic
fields are defined following~\cite{michael95}.

 We employed  two types of hadronic operator for the heavy-light mesons
- local and fuzzed. Then for the initial state of two  such mesons we
have  4 basis states. If one restricts to the operators symmetric under 
interchange, then this leaves three operators, symbolically LL, LF+FL
and FF. We then have   a $3 \times 3$ matrix of correlations at time $t$
between  these states at initial and final time. From this we use a
variational approach to extract  the linear combination of operators
which maximises the ground state contribution.

\section{Results}

 Our results are for quenched lattices at $\beta=5.7$ and we set the
 scale~\cite{mic:98}  from (string tension)$^{1/2}$=0.44 GeV (which
implies  $r_0=0.53$ fm) using $r_0/a=2.94$  to obtain $1/a=1.10$ GeV.
This scale has systematic errors  of at least 10\% coming from the
differences relative to experiment of  different observables in the
quenched approximation. There will also be  lattice corrections which
should be dominantly of order $a^2$ since we use  clover improvement.
Because of the similarity  with the lattice spacing and GeV units, we
present most of our results  in lattice units with the understanding
that they can be read as GeV to  get an estimate of the physical units.
Thus we are able to measure  the strength of the interaction  out to 
separations of  $R \approx 8$ which will correspond roughly to 1.4 fm.

 For the ${\cal B}$ meson itself, needed to evaluate binding energies,
we follow Ref.~\cite{mic:98} and use either variational analyses or a 
fit to all correlations over a range of $t$-values. We find that there
are  substantial excited state contributions and that a  good two-state
fit is possible to our correlations from 60 gauge configurations for  $5
\le t$ with $\chi^2/{\rm dof} = 2.4/(15-6)$  yielding $m_B=0.876(6)$.
This can be contrasted with the value of 0.875(6) obtained  in
Ref.~\cite{mic:98} from a fit for $5 \le t $ to a larger  variational
basis from 20 gauge configurations. For a variational study, we
determine the  basis from using $t$ of 3 and 4 and then follow the
effective mass in that  basis to larger $t$ to look for a plateau which
we find by $t$ values of 6 and 7 -- see Table~\ref{bb0}. This  gives
similar results  to the fit approach.

 For a study of the ${\cal B}$${\cal B}$ system, one approach would be
to use the variational  basis found in the ${\cal B}$ meson study for
each of the two ${\cal B}$ mesons. This will  certainly be a good
approach at large $R$ when any interaction between  the two ${\cal B}$
mesons will be very small. We shall use this basis to  give an overview
of the relative size of different contributions to  the interaction.

 A more sophisticated approach would be to make a new variational study
of the  ${\cal B}$${\cal B}$ system itself. The spatially-symmetric
sector is described by a $3 \times 3$ matrix  as discussed above. We
find in practice that this  ${\cal B}$${\cal B}$ optimal basis gives
very similar results to using the ${\cal B}$ meson basis  for each
${\cal B}$ meson.

 Given that a combined fit was found to be the method of choice for the 
${\cal B}$ meson study~\cite{mic:98}, we should also investigate fits to
the ${\cal B}{\cal B}$ correlations. One problem is that if the ${\cal
B}$ is described by  two states,  then ${\cal B}$${\cal B}$ will require
three energy eigenstates (${\cal B}{\cal B}$, ${\cal B}'{\cal B}$,
and ${\cal B}'{\cal B}'$). This increase in parameters makes the fit
less stable. 

In each case, we can use a bootstrap method  to study the binding energy
by using the same subsets of gauges for the  ${\cal BB}$ and ${\cal B}$
studies.

For this study, we use on-axis separations $R=0,1,\ldots,5$ for spatial 
size $12^3$ and $R=0,1,\ldots,8$ for $16^3$. We also measured the
correlation for the off-axis separation of $R=(\pm1,\pm1,0)$ in both
cases.

\begin{figure}[tb]
\vspace{11cm} 
\includegraphics{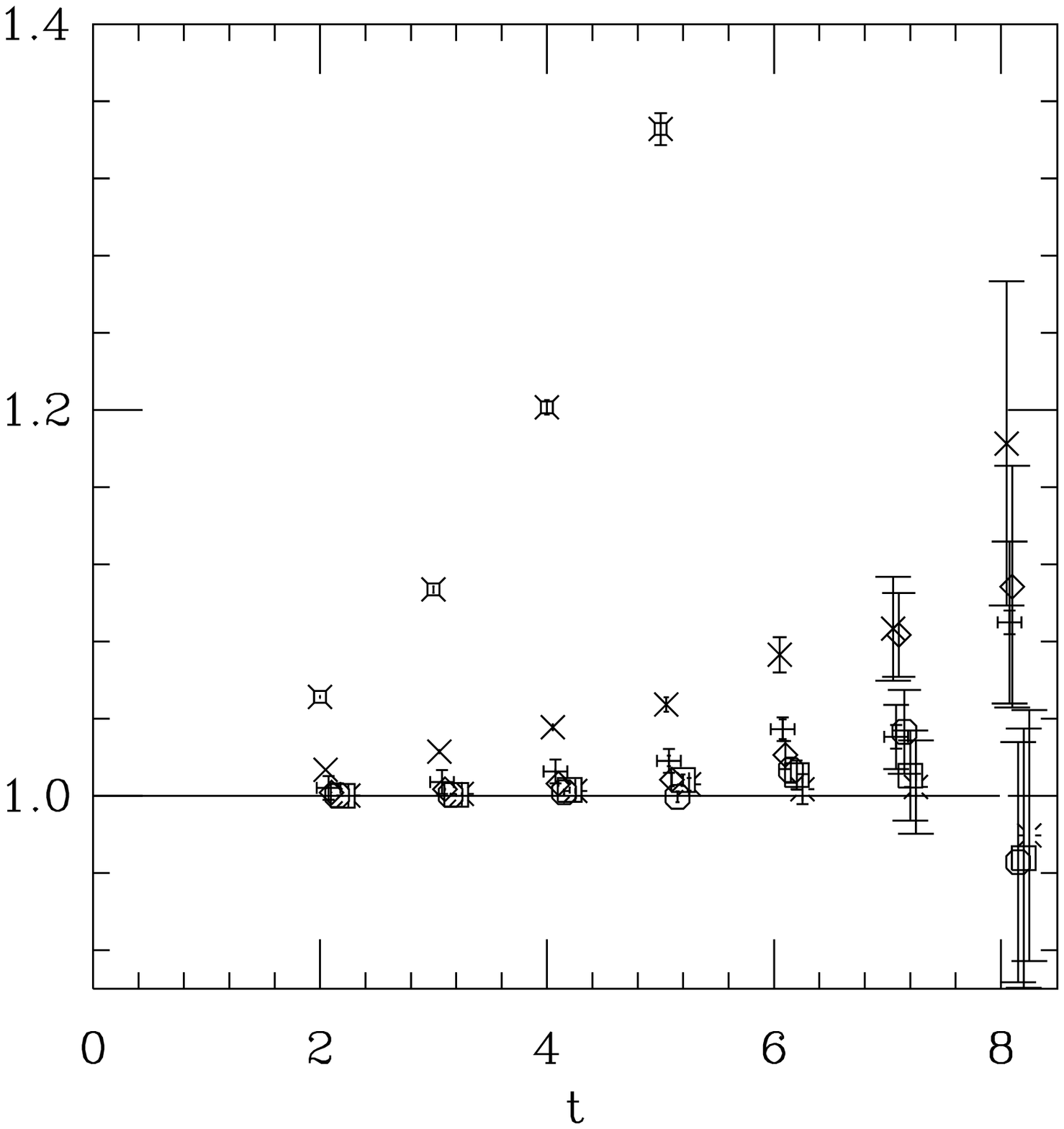}
 \caption{Results for ratio of  uncrossed diagram for the spin average
(corresponding to $C_I$) for  two ${\cal B}$ mesons divided by  the
product of two ${\cal B}$ meson correlators, versus $t$.  The separation
$R$ is 0 (fancy square),\ 1 ($\times$),\  (1,1,0) (fancy plus),\ 2
(diamond),\ 3(octagon),\ 4 (square) and 5 (*) in lattice units.
 }
 \label{bbra}
\end{figure}

 {\em Overview of results.} We first discuss  our results from $12^3$
spatial lattices in terms of  the ratios of contributions to the
uncrossed diagram for  spin average ($C_I$), taking the  ${\cal B}$
meson basis discussed above. For the ${\cal B}$${\cal B}$ correlator in
this basis divided by the  square of the ${\cal B}$ correlator in the
same basis, we find the results given  in Fig.~\ref{bbra}. This shows
that, for $R > 2$, we find this ratio  to be consistent with constant
versus $t$. This constancy implies that there  would be no binding
energy for this correlation within the errors.  The fact that the ratio
is larger than one can be explained as a  consequence of the fluctuation
with spatial location and gauge configuration  of the ${\cal B}$ meson
correlator and the  property that $\langle c^2 \rangle > \langle c
\rangle^2$ for a fluctuating quantity $c$. The relatively large
statistical  error on the signal of interest, the departure of the ratio
in Fig.~\ref{bbra} from 1.0,  is consistent with the observation that
disconnected quark diagrams  (as this is) are noisier in lattice
simulations than connected diagrams (such  as the crossed diagram). 

\begin{figure}[tb]
\vspace{11cm} 
\includegraphics{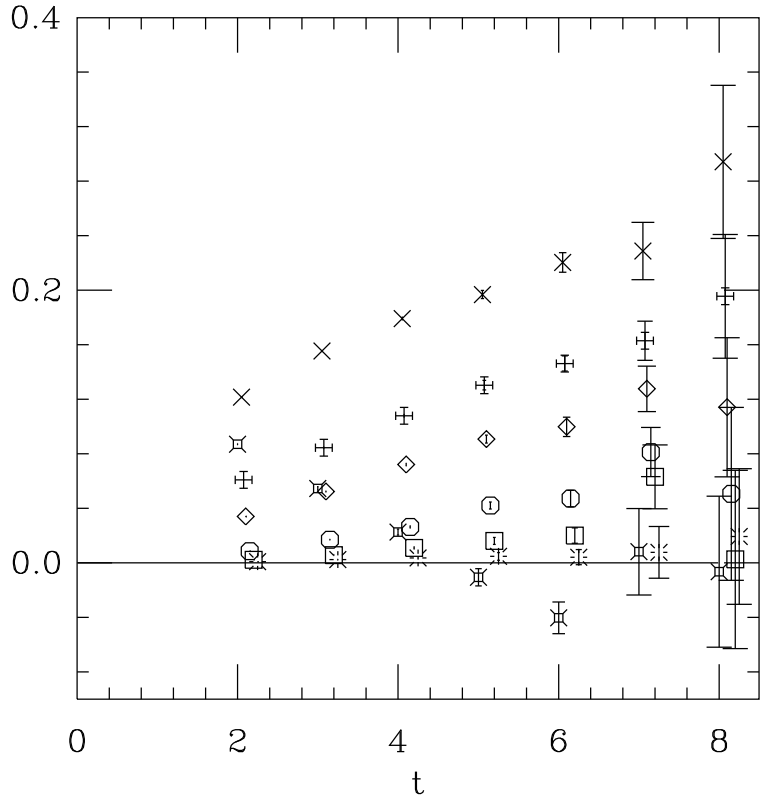}
 \caption{Results for ratio of cross diagram  to uncrossed diagram for
the spin average (corresponding to $C_I$) for  two ${\cal B}$ mesons 
versus $t$.
 The separation $R$ is 0 (fancy square),\ 1 ($\times$),\  (1,1,0) (fancy
plus),\ 2 (diamond),\ 3(octagon),\ 4 (square) and 5 (*) in lattice
units.
 }
 \label{bbcr}
\end{figure}

 The ratio of the spin-averaged ${\cal B}$${\cal B}$ correlation from
the cross diagram to that from the uncrossed diagram is shown in
Fig.~\ref{bbcr}. The ratio  is seen to increase with $t$ and to decrease
with $R$ ($R=0$ is anomalous). This $t$-dependence implies an
interaction, and we find it to decrease  in relative strength with
increasing $R$.

\begin{figure}[tb]
\vspace{11cm} 
\includegraphics{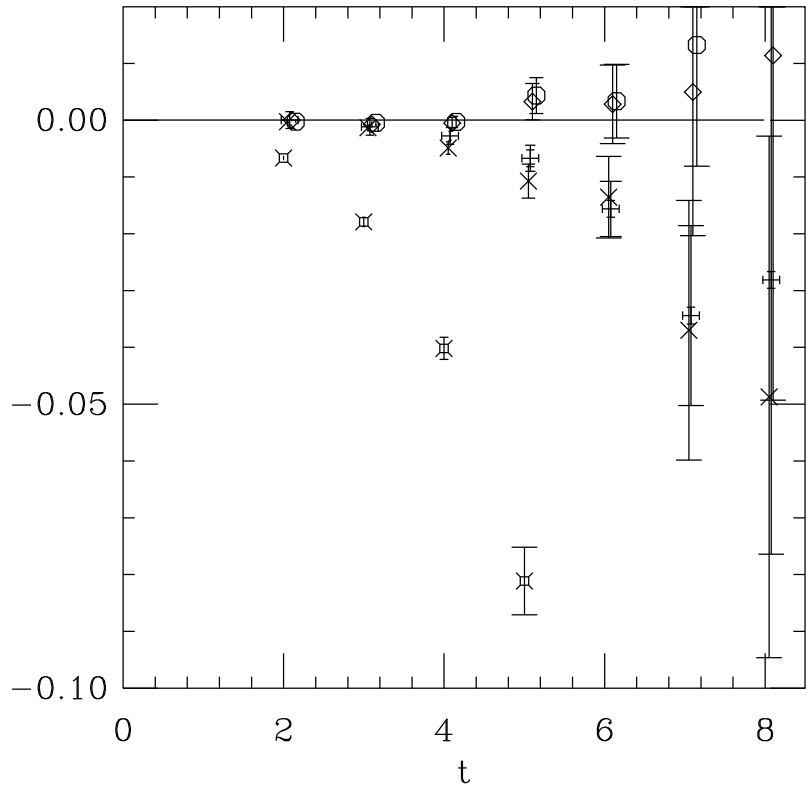}
 \caption{Results for ratio of spinflip to spin average (corresponding
to $C_s/C_I$) for  the uncrossed diagram with two ${\cal B}$ mesons
versus $t$. The separation $R$ is 0 (fancy square),\ 1 ($\times$),\ 
(1,1,0) (fancy plus),\ 2 (diamond) and 3(octagon) in lattice units.
 }
 \label{bbsp}
\end{figure}

\begin{figure}[tb]
\vspace{11cm} 
\includegraphics{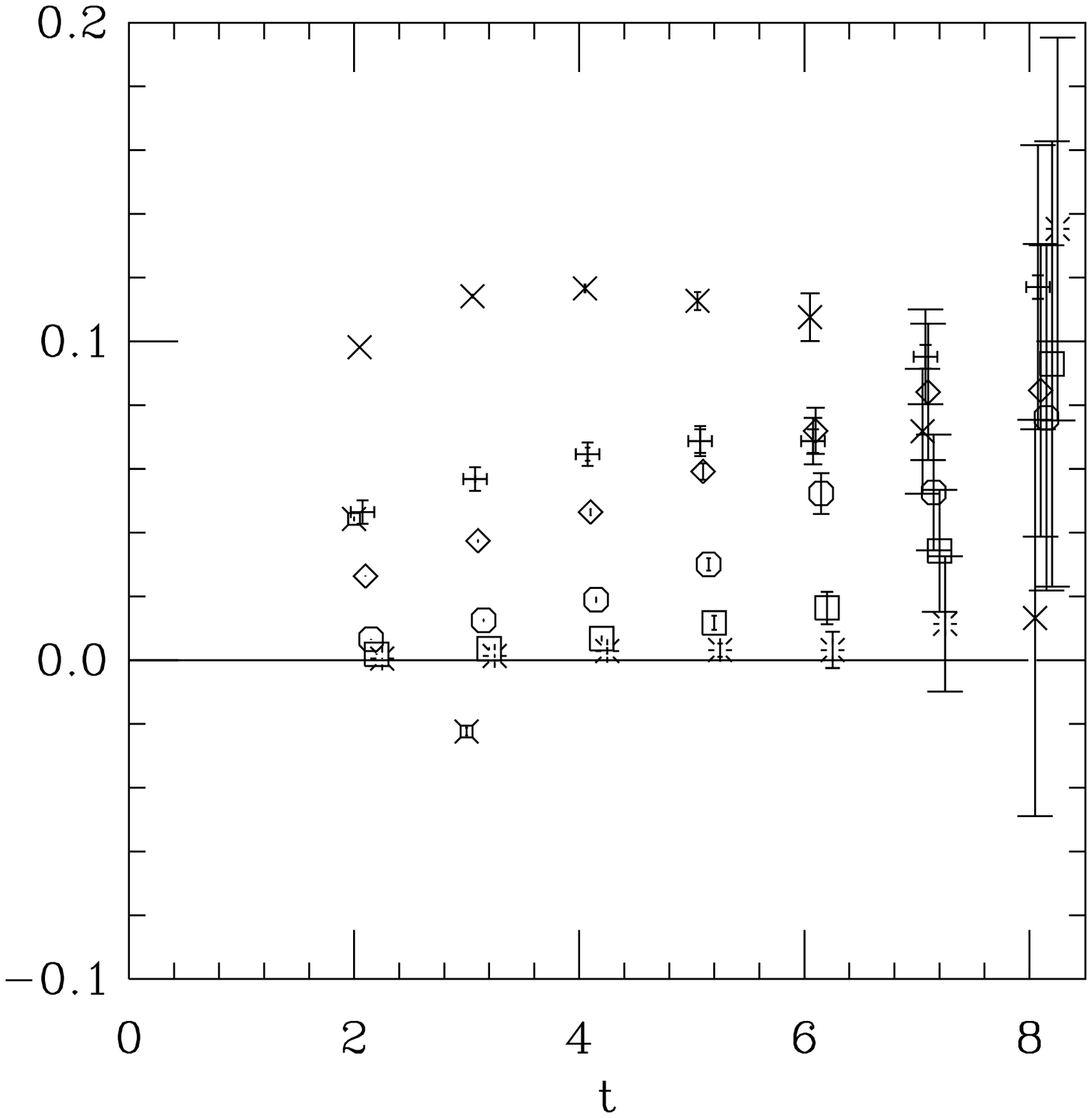}
 \caption{Results for ratio of spinflip cross diagram to spin average 
uncrossed diagram (corresponding to $C_s/C_I$) for  two ${\cal B}$
mesons versus $t$. The separation $R$ is 0 (fancy square),\ 1
($\times$),\  (1,1,0) (fancy plus),\ 2 (diamond),\ 3(octagon),\ 4
(square) and 5 (*) in lattice units.
 }
 \label{bbcrfl}
\end{figure}

The uncrossed spin flip correlation ($C_s$ averaged over $x$,\ $y$ and
$z$) is  fairly small as shown in Fig.~\ref{bbsp} and has big
errors. The dominant contribution  to the spin-flip comes from the cross
diagram as illustrated in Fig.~\ref{bbcrfl}. In both cases, the spin-flip 
correlation is poorly determined at larger $R$. We shall discuss these 
contributions in terms of particle exchange later.

We find that  the uncrossed diagram  mainly contributes to the spin
average, while the crossed diagram contributes a comparable amount to
the  spin-flip and spin average.  This is easy to understand since
crossing the  quarks will cause the spin average component ($s_1 s_2 \to
s_3 s_4$ of $+- \to +-$) to become  $+- \to -+$ which is spin flip.  We 
also looked at the tensor interaction ($2C_s(z)-C_s(x)-C_s(y)$) but 
found a small and poorly determined signal.

In the analysis presented above, the ${\cal B}$ meson ground state has
been extracted by using the variational basis found from a study of a
single ${\cal B}$ meson. It  is not feasible to construct a pure two
meson state on a lattice in Euclidean time  since asymptotic states
cannot be constructed. Rather, one can only  construct a state with
given quantum number and then extract the  energy eigenvalues.
Nevertheless, a qualitative understanding can be obtained,  as above,
by constructing approximations to the two meson state  and exploring 
their correlations.  

 For example if we find that the ratio 
 $C_s/C_I \approx C +Dt$,
 and  if the combinations $C_I+f_1C_s$ and $C_I+f_2C_s$ correspond to
two given sets of  quantum numbers 1, 2 in the ${\cal BB}$ channel, then
the mass difference  which is obtained from the $t$-dependence of these
correlators at large $t$ satisfies 
 \be
E_1-E_2= {d \over dt} \log{C_I+f_2 C_s \over C_I+f_1 C_s} \approx
{d \over dt} (f_2-f_1)(C+Dt) \approx (f_2-f_1)D
 \ee
 if $C_s/C_I$ is small.  Thus a linear dependence of the  ratio is
expected and can be related to the energy difference as shown.  We do
indeed see evidence for such linear behaviours  of ratios in the figures
just discussed.   
 This linear dependence of the spin-flip correlation on $t$ can also be
related theoretically to a meson exchange interpretation, for example,  
and we discuss this later.

{\em Energy Levels.} A study of correlations between lattice operators
at increasing  $t$ allows an analysis of energy levels. Thus by taking
the appropriate  combination of crossed and spin-flip contributions, the
energy of ${\cal B}$${\cal B}$ states  with different quantum numbers
can be studied. Because the binding energies  are found to be quite
small, we have presented the foregoing discussion to  show the quality
of our lattice data.

\begin{figure}[tb]
\vspace{11cm} 
\includegraphics{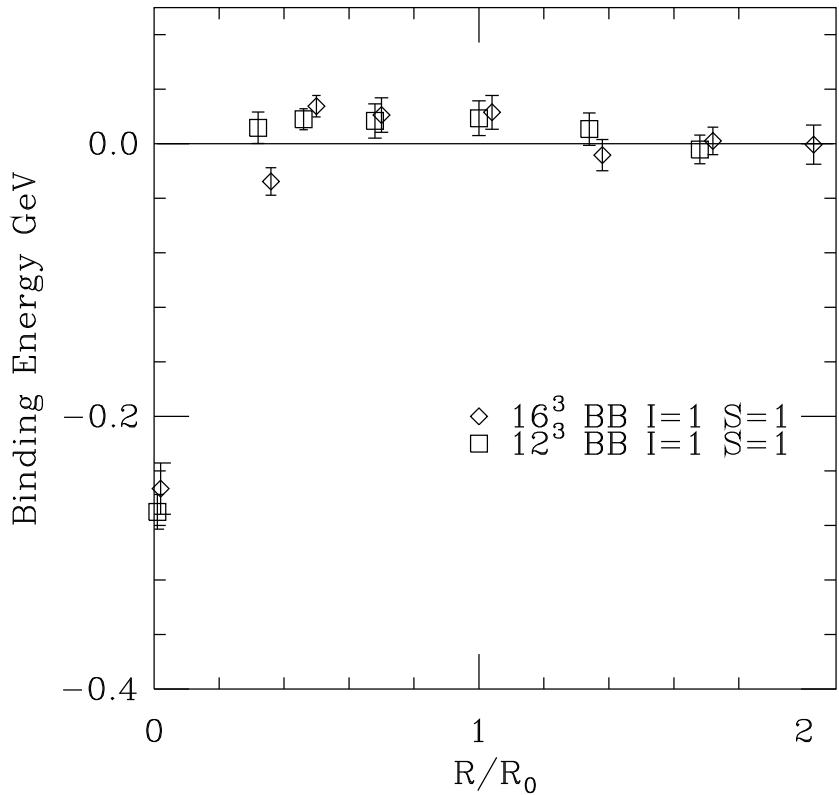}
 \caption{Results for the binding energy  between two ${\cal B}$ mesons with 
light quarks in $(I_q,S_q)$=(1,1)  at separation $R$
in units of $R_0 \approx 0.5$fm. The light quark mass used corresponds
to strange quarks.  Results from  variational method using basis from
$t$ 4:3  and effective mass  in that basis from $t$ 6:5. Results at 
different spatial lattice sizes are displaced in $R$ for legibility. 
 }

 \label{bb411}
\end{figure}

\begin{figure}[tb]
\vspace{11cm} 
\includegraphics{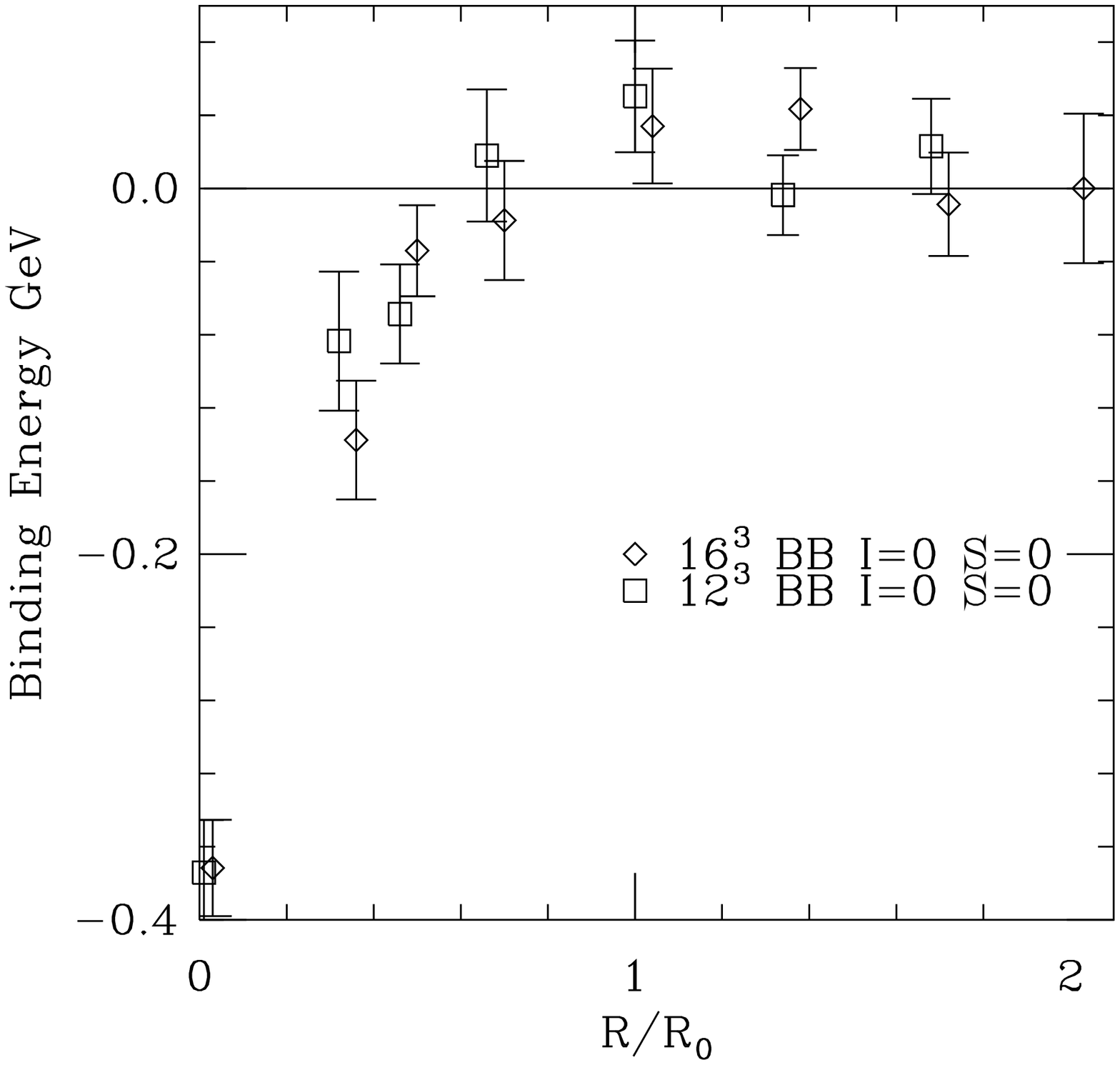}
 \caption{Results for the binding energy  between two ${\cal B}$ mesons with 
light quarks in $(I_q,S_q)$=(0,0) at separation $R$
in units of $R_0 \approx 0.5$fm. The light quark mass used
corresponds to strange quarks.  Results from  variational method using
basis from $t$ 4:3  and effective mass  in that basis from $t$ 6:5.
 }
 \label{bb400}
\end{figure}

\begin{figure}[tb]
\vspace{11cm} 
\includegraphics{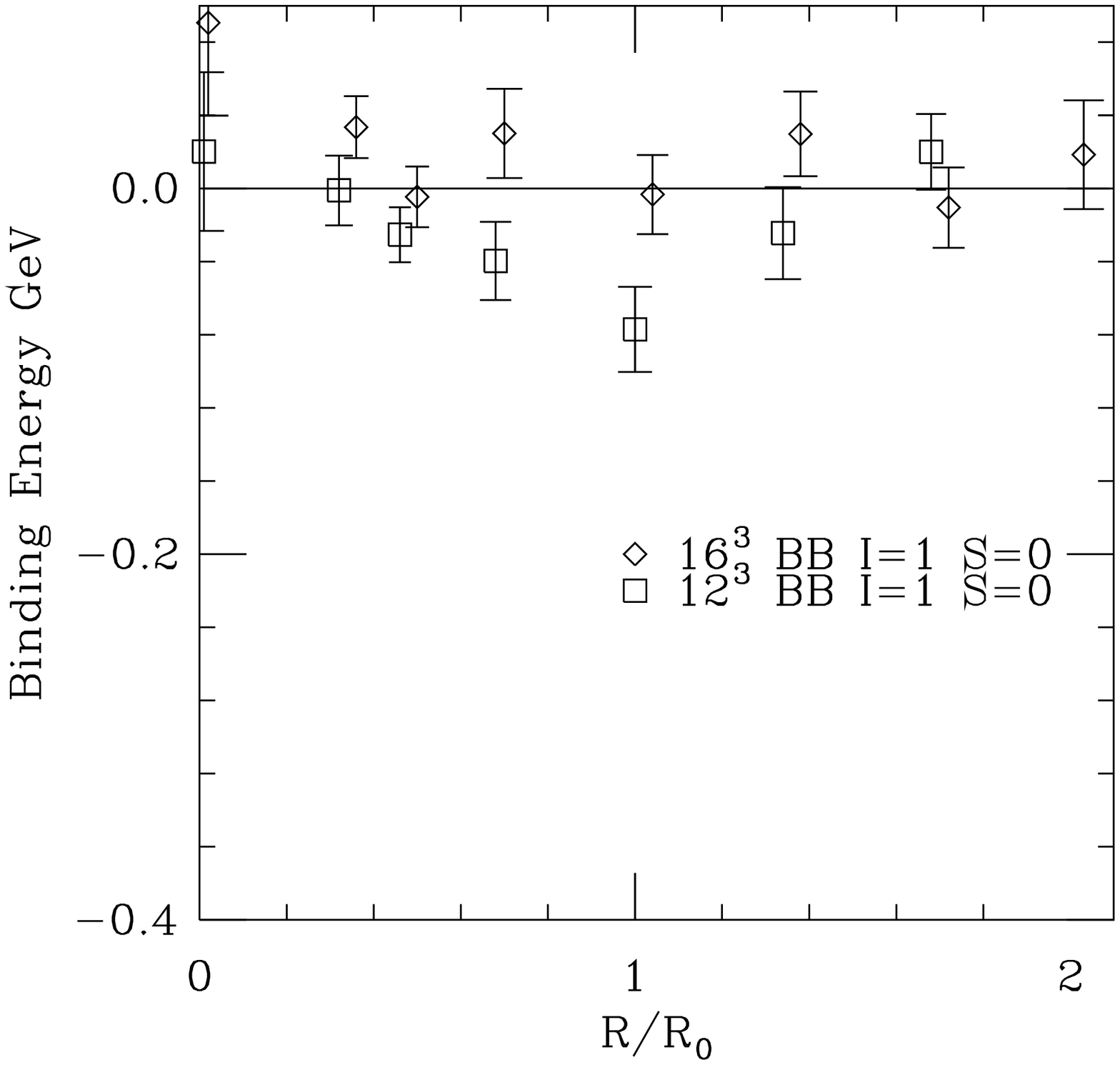}
 \caption{Results for the binding energy  between two ${\cal B}$ mesons with 
light quarks in $(I_q,S_q)$=(1,0) at separation $R$
in units of $R_0 \approx 0.5$fm. The light quark mass used
corresponds to strange quarks.  Results from  variational method using
basis from $t$ 4:3  and effective mass  in that basis from $t$ 6:5.
 }
 \label{bb410}
\end{figure}

\begin{figure}[tb]
\vspace{11cm} 
\includegraphics{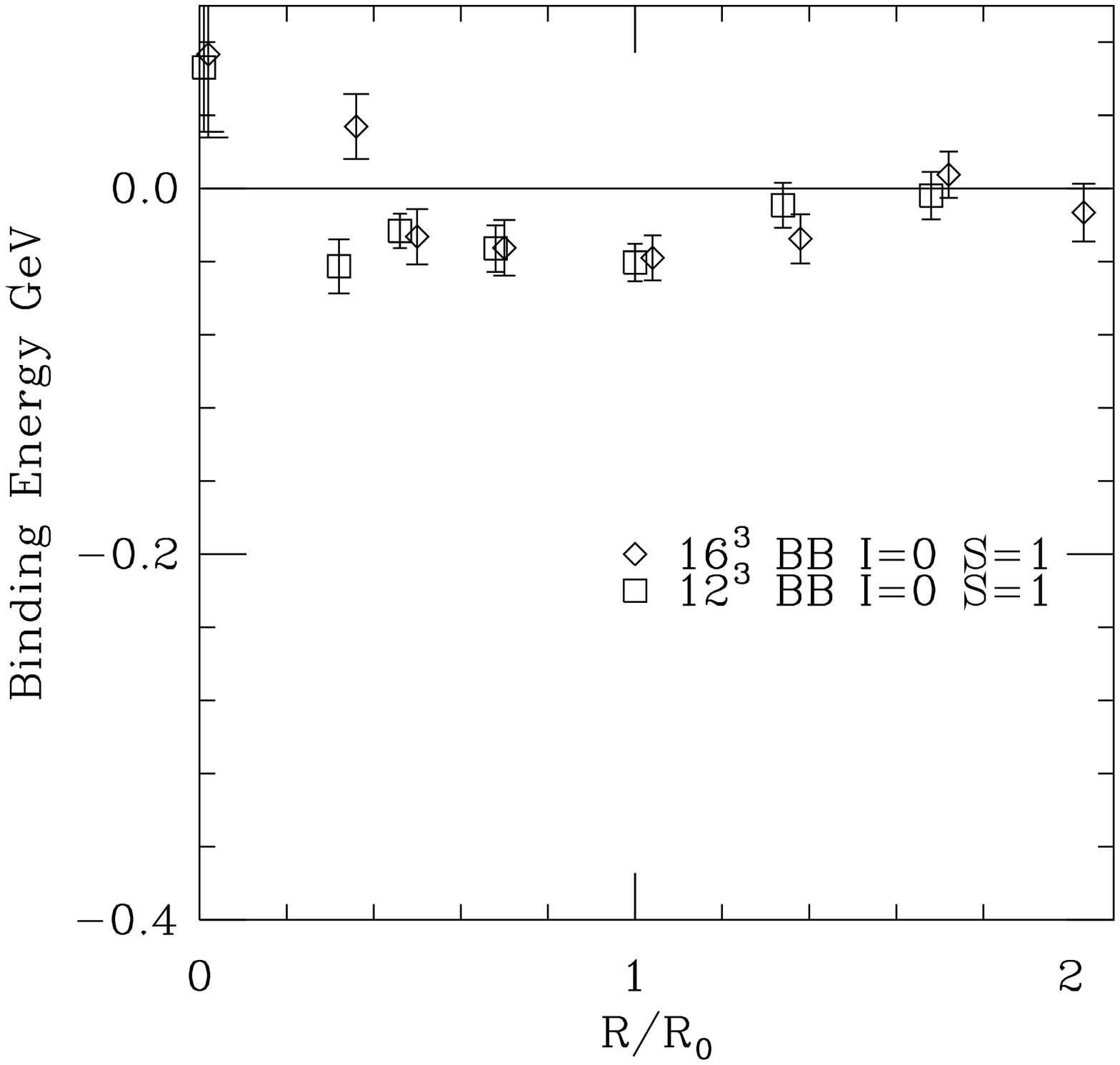}
 \caption{Results for the binding energy  between two ${\cal B}$ mesons with 
light quarks in $(I_q,S_q)$=(0,1) at separation $R$
in units of $R_0 \approx 0.5$fm. The light quark mass used
corresponds to strange quarks.  Results from  variational method using
basis from $t$ 4:3  and effective mass  in that basis from $t$ 6:5.
 }
 \label{bb401}
\end{figure}

We  present the  energies for isospin 0 and 1 light quarks for the
triplet and singlet spin combinations,  using a $3\times 3$ variational
basis from $t$ of 4 and 3 to obtain the optimum  combination for the
${\cal B}$${\cal B}$ ground state.   Here we use $C_I+C_s$ correlation
for the triplet states with $C_s$ the average over orientation which is 
appropriate for an $S$-wave bound state and  $C_I-3C_s$ for the singlet 
states.  The energies evaluated on a lattice include a contribution from
the self-energy of the static source which is unphysical. Thus only 
energy differences have a physical significance and hence we 
concentrate especially on the binding energies -- the difference of
${\cal BB}$  energy from twice the ${\cal B}$ energy. In the special 
case of $R=0$, we show the actual lattice energy values in 
Table~\ref{bb0} to allow us to discuss the extrapolation to large  $t$
needed to extract the ground state. Other results  are  given in 
Table~\ref{bbr} for the  case of $R=3$ and  in Figs.~\ref{bb411} to
\ref{bb401}.  We show the results from both $12^3$ and $16^3$ spatial
lattices  with the same parameters in order to explore finite size
effects. Within errors, we do not see significant differences in the 
results between spatial sizes of $L=12$ and 16, which is not  unexpected
since a study of the ${\cal B}$ meson using $L=8$ and 12 
found~\cite{mic:98} agreement for the energies of the ground state
mesons and  a  relatively localised Bethe-Saltpeter wave function.

\begin{table}[hbt]

\caption{Effective masses for ${\cal B}$ and for ${\cal BB}$ at R=0}
\begin{center}
\begin{tabular}{cccllll}
% In 4/3 variational basis  
 $  I_q $ &  $S_q$ & $L$ & \multicolumn{4}{c}{$E_{\rm eff}$} \\
&&&\multicolumn{3}{c}{$t$-ratio:} & ref{\cite{mic:98}}\\
&& &  5/4 &  6/5  & 7/6 & \\
\hline
 ${\cal B}$ &&&&&&\\
    1/2   & 1/2  &12    &0.911(3)  &0.893(3)  &0.873(6)    &0.875(6)\\
 ${\cal BB}$&&&&&&\\
    1     & 1    &12    &1.620(9)  &1.516(12) &1.558(32)   &1.514(52)\\
          &      &16    &1.589(10) &1.539(18) &1.476(35) &\\
    0     & 0    &12    &1.472(18) &1.412(29) &1.301(63)   &1.435(37)\\
          &      &16    &1.458(17) &1.420(31) &1.348(48) &\\
    1     & 0    &12    &1.864(22) &1.806(44) &1.629(111)   &\\
          &      &16    &1.915(19) &1.882(50) &1.827(130)   &\\
    0     & 1    &12    &1.911(19) &1.852(36) &1.722(109)   &\\
          &      &16    &1.860(18) &1.865(43) &2.886(134)   &\\

\end{tabular}
 \label{bb0}
\end{center}
\end{table}

\begin{table}[hbt]

\caption{Binding energies for ${\cal BB}$ at R=3a}
\begin{center}
\begin{tabular}{cccccc}
% In 4/3 variational basis  
 $  I_q $ &  $S_q$ & $L$ & \multicolumn{3}{c}{$E({\cal BB})-2 E({\cal B})$}\\
&&&\multicolumn{3}{c}{$t$-ratio:} \\
&& &  5/4 &  6/5  & 7/6  \\
\hline
    1     & 1    &12    &0.027(5)  &0.019(13) &0.012(32) \\
          &      &16    &0.024(4)  &0.023(12) &0.005(41)\\
    0     & 0    &12    &0.035(10) &0.050(30) &0.021(90)\\
          &      &16    &-0.013(13)&0.034(32) &-0.000(85)\\
    1     & 0    &12    &-0.002(10)&-0.077(23)&0.040(72)\\
          &      &16    &-0.030(10)&-0.003(22)&0.069(92)\\
    0     & 1    &12    &-0.029(6)&-0.040(10)&-0.061(38)\\
          &      &16    &-0.021(7) &-0.038(12)&-0.019(42)\\

\end{tabular}
 \label{bbr}
\end{center}
\end{table}

 The situation at $R=0$ is special because the two static $b$ quarks 
can be classified under their combined colour into either an
anti-triplet or a sextet. The former case is just that which applies to
baryons with one  static quark and these are expected to be the lightest
states. Thus the ${\cal BB}$ spectrum at $R=0$ is expected to reproduce
these  baryonic levels.   As shown in  Table~\ref{bb0}, we find
excellent agreement with the masses of  baryonic states with one static
quark which have been obtained  on the lattice previously~\cite{mic:98}.
This is a useful cross check of our  procedures for obtaining energy
levels. Thus we find that the  $\Lambda_b$  (with light quarks of $I=0$
and in a spin singlet) is the lightest state. Combining  with the ${\cal
B}$ meson mass then gives  an estimate of the binding energies at $R=0$
which will agree well  with those we obtain here -- namely around 400 MeV 
for the $I_q,S_q=(0,0)$ state. 
  We are also able to explore the  energies of states with $R=0$ having
the opposite symmetry -- so corresponding to the sextet of colour which
is symmetric (rather than the anti-triplet which is antisymmetric). We
find these states to lie  higher in mass than the anti-triplet states by
about 0.3 in lattice units  as shown in Table~\ref{bb0} and to be
unbound.

 Unlike on the lattice, where for static quarks, the binding energy at
$R=0$ can be  obtained by taking the difference of the lattice baryon
mass with twice  the lattice ${\cal B}$ mass, 
 in the continuum, in the heavy quark limit, one would expect that the
binding of the ${\cal BB}$ system at $R=0$  for light quark
flavour of $I=0$ is given by  $2(M_B-m_b)-(M_{\Lambda_b}-m_b)$ where
$m_b$ is the $b$-quark mass.

 Since we find that the variational method gives a plateau from 
$t$-values  of 5 and 6, as shown in Table~\ref{bb0}, we expect that that
would be a good criterion to use for $R >0$.

However, for the ${\cal B}$ meson case itself, it is found that our
variational method  does not achieve  a plateau value for the effective
mass until  a $t$-ratio of 7 to 6 as shown in Table~\ref{bb0}. At these
large $t$-values,  the ${\cal BB}$ signal is very noisy. Since the same
operators are used for  the ${\cal B}$${\cal B}$ case as for the ${\cal
B}$ meson alone, it is feasible that excited state  contributions are
dealt with similarly in each case, particularly  for $R >0$ where the
binding energy is found to be very small. Thus it makes sense to  study
the difference (the binding energy) obtained from the ${\cal BB}$
effective mass at a given $t$-ratio and twice the ${\cal B}$ meson
effective mass at the same $t$-ratio. This is plotted in
Figs.~\ref{bb411} to  \ref{bb401} from the ratio of correlations at 
$t$-values of 5 and 6.

To explore the consistency of the binding energy obtained in this  way,
we show in Table~\ref{bbr} at $R=3$ the variational effective mass
differences  from different $t$-values.  This leads us to conclude that
the variational effective mass values  for the binding energy are
consistent with being constant within  errors from $t$-values of  5/4, the
excited state contamination being smaller than for the total energies. 
For extra safety in extracting the ground state,  we shall use the
effective mass from the $t$-values of  6/5, as stated above.  

As a cross check of this procedure, we find that the binding energy is 
consistent with zero within errors at large $R$, namely $R \ge 5$.

 As one goes to nonzero $R$, the level ordering found at $R=0$ would be
expected to be retained if the dominant dynamical configuration was 
that the two heavy quarks combine to an anti-triplet.  We illustrate the
binding energies for these states analogous to the $\Lambda_b$ in
Fig.~\ref{bb400} and the $\Sigma_b$  in Fig.~\ref{bb411}.  We see the
level ordering to persist for the smallest values of $R$, the binding
disappearing at $R\approx 0.2$ fm for the $\Sigma_b$-like state and at
$R\approx 0.3$ fm for the $\Lambda_b$ analogue.  The binding for the
other pair of states is shown in  Figs.~\ref{bb410} and \ref{bb401}.  
Here we see that the $I_q,S_q$=(0,1) state shows a statistically
significant binding of 40 MeV at  $R \approx 0.5$ fm. The situation  for
the $I_q,S_q$=(1,0) state is less clear, since the statistical 
fluctuation is larger, but it is consistent with a similar
interpretation. Note that  pion exchange in the cross diagram will act
to make the $I_q,S_q$ = (1,0)  and (0,1) states lightest at large $R$ as
we discuss in more  detail later.

\section{Discussion}

{\em Bound states}.  We find binding at small $R$ for $I_q,S_q$=(0,0) and
(1,1) and  binding at moderate $R$ (circa 0.5 fm) for (1,0) and (0,1). 
For very  heavy quarks, this will imply binding of the ${\cal BB}$
molecules with these quantum numbers and $L=0$. For the physically
relevant case  of $b$ quarks of around 5 GeV, the kinetic energy will
not be negligible  and the binding energy of the ${\cal BB}$ molecular
states is less  clear cut. One way to estimate the kinetic energy for
the ${\cal BB}$ case with reduced mass circa 2.5 GeV is to use analytic
approximations to the  potentials we find. For example the
$I_q,S_q$=(0,0) case shows a deep  binding at $R=0$ which we can
approximate as a Coulomb potential of $-0.1/R$ in GeV units. This will
give a di-meson binding energy of only 10 MeV.  For the other
interesting case, $I_q,S_q$=(0,1), a  harmonic oscillator potential in
the radial coordinate of form $-0.04[ 1- (r-3)^2/4]$ in GeV units leads
to a kinetic energy  which completely cancels the potential energy
minimum, leaving zero  binding. This harmonic oscillator approximation
lies above our estimate of  the potential, so again we expect weak
binding of the di-meson system.

 Because of these very small values for the di-meson binding energies, 
we need to retain corrections to the heavy quark approximation to 
make more definite predictions, since these corrections are known to 
be of magnitude 46 MeV for the ${\cal B}$ system. It will also be 
necessary to extrapolate our light quark mass from strange to 
the lighter $u,\ d$ values to make more definite predictions 
about the binding of B mesons. This is especially necessary for light 
meson exchange contributions, which  discuss subsequently.

A model for static four-quark systems is extended and fitted to
our binding energies in Ref.~\cite{gre:99}. As in the static case, the
results point out the inadequacy of a  simple two-body potential
approach for describing multi-quark systems. Inclusion of a multi-quark
interaction term interpolating between strong and weak coupling regimes
enables  reproduction of the lattice data.

{\em One meson exchange}. The interaction responsible for the binding
energy in the ${\cal BB}$ system  can be discussed in terms of meson
exchange. One simple criterion  is that BB\  $\to$\  BB only allows
natural parity exchange (such as vector meson exchange) while  BB$^*$\
$\to$\ B$^*$B has an unnatural parity exchange component as well. Here
natural means that the exchanged mesons have parity $(-1)^J$.  This can
be explored by viewing the diagrams of Fig.~1 as representing a
(spatially non-local) meson creation at $z=z_1$ and then annihilation at
$z_2=z_1+R$. The quantum numbers of the mesons propagating in the 
$z$-direction then can be determined from the Dirac structure of the 
effective creation operator. So for $C_I$ (BB\ $\to$ BB), we have scalar
and  vector mesons allowed (natural parity exchanges), while for
$C_s(z)$, we have pseudoscalar  and axial (unnatural parity exchanges),
while for $C_s(x)$ and $C_s(y)$, both axial and vector 
are allowed. From this analysis it follows that at large
$R$, the correlations at fixed $t$  behave as $\exp(-MR)$ with $M$ the
mass corresponding to the lightest meson exchange allowed.   For our
lattice parameters, these will be the pseudoscalar meson, mass 0.529(2),
for $C_s(z)$ and vector meson, mass 0.815(5), for $C_I$. 
 
  Meson exchange contributes to   the uncrossed diagram with flavour
singlet exchange only while the  crossed diagram has both  flavour
singlet and   non-singlet mesons exchanged. In the quenched
approximation, the flavour singlet and flavour non-singlet mesons are
degenerate. However, in full QCD,  the flavour singlet mass is  modified
 by quark loop effects which are not present in the quenched case. These
effects are responsible for the $\eta$, $\eta'$ mass splitting, for
example. Thus to make the cleanest comparison with meson  exchange, it
is appropriate to use the flavour non-singlet mesons  ($\pi$, $\rho$
etc) which contribute only to the crossed diagram.  Furthermore, our
determinations of the contributions from the uncrossed diagram are
considerably more noisy, so this comparison with the crossed diagram 
alone will be  a tighter test. 

\begin{figure}[tb]
\vspace{11cm} 
\includegraphics{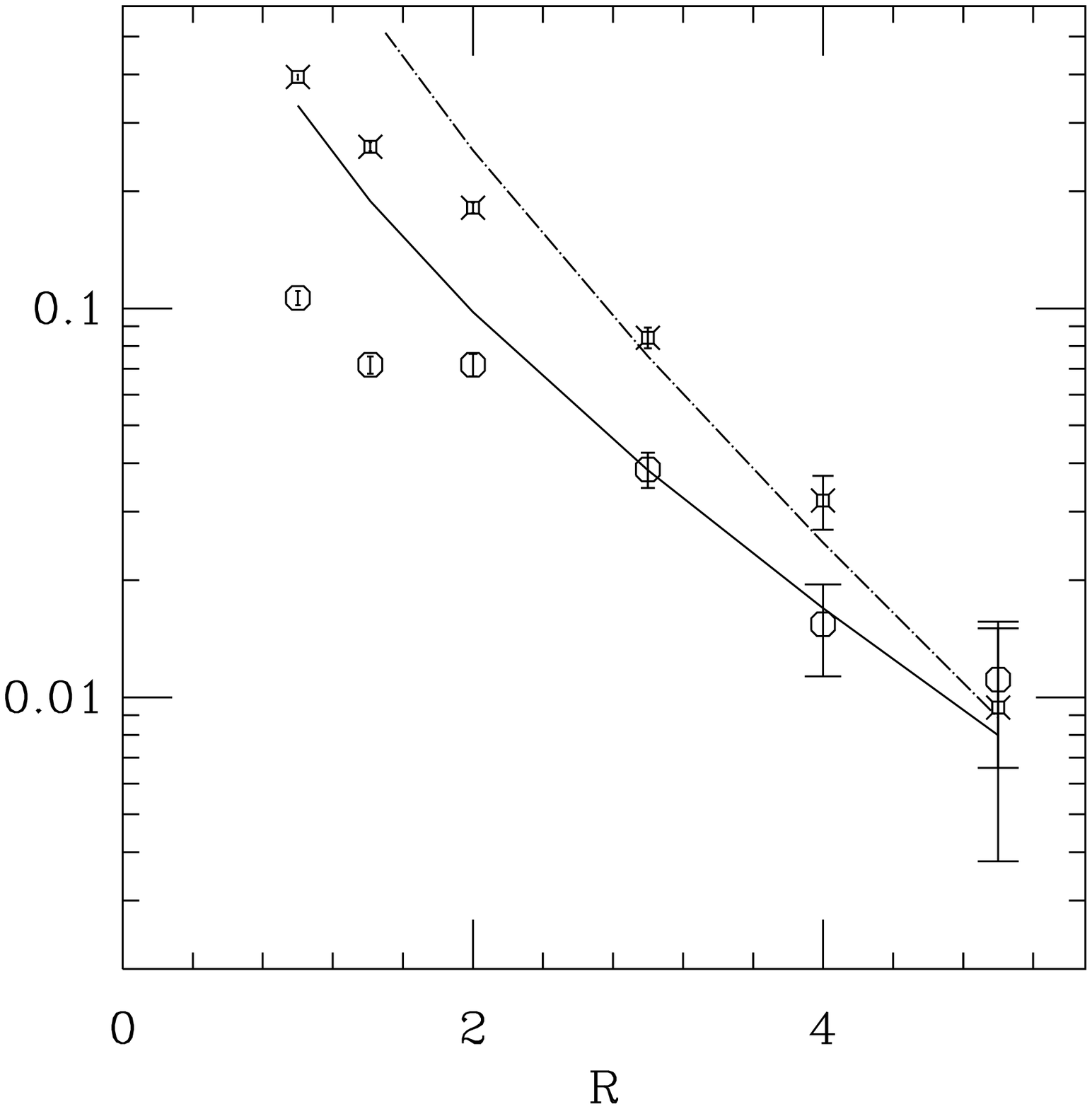}
 \caption{The ratio of the crossed-diagram contributions to  the spin
averaged uncrossed contribution for the ${\cal B B}$ correlation at
$t$=5. Shown are the  crossed diagram correlation for the  spin average
(BB\ $\to$\ BB, $C_I$, multiplied by two for clarity of presentation,
fancy squares) and the spin-flip (BB$^*$\ $\to$\ B$^*$B, $C_s(z)$,
octagons). The meson exchange expressions,\ $\exp(-MR)/R$,  are compared
with these results for  $C_s(z)$ (using pion exchange with $M=0.529$, 
continuous line) and  for $C_I$ (rho  exchange with $M=0.815$, dotted).
Note that the pion exchange expression is normalised as described in the
text, whereas the rho exchange contribution has an ad hoc normalisation.
 }
 \label{pirho}
\end{figure}

 Then, as shown in Fig.~\ref{pirho}, we  see evidence for an exponential
decrease of the interaction with increasing  separation $R$ with a mass
exponent consistent with that expected, namely,  vector for $C_I$ and
pseudoscalar for $C_s(z)$. This agreement with the nature of the lightest 
meson exchange is a confirmation that the arguments given above apply 
at modest $R$ values. 
  Since the lattice operator which creates the  meson is  not at zero
momentum, we expect non-exponential  contributions to yield  the
expression $(1/R)\exp(-MR)$ where we have assumed that a sum over the
$t$-direction is taken (so $t$ is large: here we need $t > (2R/M)^{1/2}$
which is satisfied in our case). This expression is just the
conventional  Yukawa potential. 

 It is possible to go further, since lattice estimates for the
B$^*$B$\pi$ coupling  are available~\cite{bbpi} from a study of the
axial  matrix element between B and B$^*$. Indeed, as well as the
coupling itself,  this lattice study also measures the form factor -- the
spatial distribution  of the coupling -- which is found to be quite
localised. So we are be able to evaluate the magnitude  of the pion
exchange contribution using the lattice pion -- so affording a direct
comparison.

 Now consider the  interaction potential for B$^*$B $\to$ B$^*$B  with 
B$^*$  spin polarisation in the $z$-direction, which has a one pion
exchange component  at large $R$, 
 \be
 V(R)= \vec{\tau}_1.\vec{\tau}_2 { g^2 M^2 \over 4 \pi f^2} {e^{-MR} \over R}
 \ee
 where $g/f$ is the pion coupling to quarks~\cite{tor:91} and we use the
value determined from the lattice~\cite{bbpi} of $g=0.42(8)$ and where
$f$  is the pion decay constant (132 MeV). Because we wish to  compare
with our lattice results with heavier light quarks, we use the lattice 
pion mass ($Ma=0.53$). 

 Then to compare with our best determined quantity, the ratio of the crossed
diagram contribution to $C_s(z)$ to the uncrossed contribution to $C_I$, 
we assume that the ratio is small so that a linear $t$-dependence is 
appropriate, as indeed is compatible with our lattice results in 
Fig.~\ref{bbcrfl}.  This implies that 
 \be
   {C^X_s(z) \over C^D_I} = {t \over 2}{ g^2 M^2 \over 4 \pi f^2} 
  {e^{-MR} \over R}
 \ee
 and we plot this for $t=5$ in Fig.~\ref{pirho}, using the parameters
discussed above.

 The agreement is excellent --  better than should be expected  given
that $t=5$ is used and the signal is only well measured for $R<5$. In
particular, non-leading contributions will be of order $1/(MR)$ which
 is relatively large, namely $1/(MR)=0.47$ at $R=4$ for our lattice pion
exchange, also note that  some non-relativistic treatments of pion
exchange have an explicit non-leading correction factor given by
$1+3/(MR)$. This  implies that we should not take our estimate of the
magnitude of one  pion exchange as more than a rough guide at the
$R$-values we are able  to measure.  Furthermore, for consistency, we
should use  the lattice determination of $f$ for our lattice  pion mass
(which corresponds to quarks with the strange mass), hence $f$ will be
somewhat larger (by a factor of around $f^2_K/f^2_{\pi} =1.4$). What our
comparison does show, however, is that the pion exchange  contribution
to the binding can be identified reliably for $R \approx 0.5$ fm.  This
allows the realistic pion mass to be used to give predictions for the 
physical case with more confidence because of the  agreement we find for
pions heavier than the physical case.

 Note that the pion exchange contribution is to $C_s(z)$ only which will 
contribute a large tensor interaction. Thus, much as for the case  of
deuterium, this is likely to be responsible for  mixing between S and D
wave components  in the di-meson bound states. Thus implications 
for bound states are not straightforward.

 In deuson models, the analysis of the pion exchange  contribution  to
the potential makes meson-antimeson states in most cases significantly
more bound than  meson-meson systems. The  possibility is raised in
Ref.~\cite{tor:91} that B$^*$B$^*$  states bound by pion exchange  may
exist. In such models, however, the small $R$ behaviour of 
the potential is not reliable.
 As discussed above, the most fruitful way to use our results would be 
to take our non-perturbative measurement of the  binding energy at small
$R$ and to modify our  meson exchange component at larger $R$ to have the 
lighter pion mass which is physically relevant. 

\section{Conclusions}

 We study the ${\cal BB}$ system at fixed separation $R$ using static 
$b$-quarks. We present evidence for deep binding at small $R$ with the
light quark  configuration similar to that in the $\Lambda_b$ and
$\Sigma_b$ baryons -- so that the heavy quarks are in a colour-triplet
di-quark state (and the light quarks have $I_q,S_q$=(0,0) and (1,1)
respectively). This  binding energy is 400 - 200 MeV at $R=0$ but is
very short-ranged. This  binding is essentially a gluonic effect and is
rather insensitive to the  light quark mass, as shown by studies of the
static baryons  with varying light quark masses~\cite{mic:98}. At larger
$R$, around 0.5 fm, we see evidence for  weak binding when the light
quarks are in  the $I_q,S_q$=(0,1) and (1,0)  states. This can be
related to meson exchange and we find evidence of an interaction in the
spin-dependent quark-exchange (cross) diagram which is compatible with
the theoretical contribution from pion exchange  in our study. Using
lighter, and hence more physical, light quark masses, this effect will
be modified in a predictable way, although further lattice study is
needed with  light quark masses below those we use (namely strange) to 
confirm this.
 Corrections also need to be evaluated to the heavy quark limit  for
applications to realistic $b$-quarks and we need to use smaller lattice
spacings  so reaching closer to the continuum limit, together with gauge
configurations  which have the contributions from sea quarks included.

Our results show that it is plausible that exotic $bb\bar{q}\bar{q}$
di-mesons exist  as states stable under strong interactions. With the
future  lattice developments described above, it will be possible to 
give a definite answer from first principles in QCD whether this is so.

\section{Acknowledgement}

We thank T. Barnes, A.M. Green and E. Swanson for discussions and the
Helsinki  Institute of Physics for hospitality. 
 We acknowledge the support from PPARC under grants GR/L22744 and
GR/L55056 and from the HPCI grant  GR/K41663.

\newcommand{\href}[2]{#2}\begingroup\raggedright\endgroup

\end{document}